\documentclass[prd,reprint,nofootinbib,superscriptaddress,preprintnumbers,showpacs]{revtex4-1}
\usepackage{amsfonts}
\usepackage{amssymb}
\usepackage{amsmath}
\usepackage{graphicx,color}
\usepackage[justification=raggedright]{caption} % 标题左对齐
\usepackage{dcolumn}
\usepackage{epsfig}
\usepackage{bm}
\usepackage{bbm}
\usepackage{ulem}
\usepackage{hyperref}
\usepackage{lineno}
\usepackage{float}
\usepackage{subcaption}
\usepackage{slashed}
\usepackage{CJKutf8}
\allowdisplaybreaks[4]

\begin{document}
	
	\title{Single spin asymmetry $ A _ { U L } ^ { \sin ( 3 \phi _ { h } - \phi_{ R } ) }$ in dihadron production in SIDIS}
	\author{Lei Tan}
	\affiliation{School of Physics and Optoelectronics Engineering, Anhui University, Hefei 230601, People's Republic of China }
	\author{Gang Li}
	\affiliation{School of Physics and Optoelectronics Engineering, Anhui University, Hefei 230601, People's Republic of China }
	\author{Mao Song}
	\affiliation{School of Physics and Optoelectronics Engineering, Anhui University, Hefei 230601, People's Republic of China }
	\author{Xuan Luo}	
	\email{xuanluo@ahu.edu.cn}
	\affiliation{School of Physics and Optoelectronics Engineering, Anhui University, Hefei 230601, People's Republic of China }
	\begin{abstract} 
In the field of particle physics, the phenomenon of dihadron production in semi-inclusive deep inelastic scattering (SIDIS) process has always been a significant focus. This paper focuses on the single longitudinal spin asymmetry $A_{UL }^{\sin(3\phi_{h}-\phi_{R})}$ in the dihadron production during this process and combines the transverse-momentum-dependent dihadron fragmentation function (DiFF) $H_1^{\perp}$ to deeply analyze its underlying mechanism. Here, the involved DiFF $H_1^{\perp}$ is the analogue of the Collins function for single-hadron production and it describes the fragmentation of a transversely polarized quark at leading twist. Recent studies have shown that the azimuthal asymmetry signal observed by the COMPASS collaboration in the dihadron SIDIS is weak. To reveal the reason for this small signal and to study the asymmetry, we calculate the unknown T-odd DiFF $H_1^{\perp}$ using the spectator model. The spectator model, widely used in SIDIS, describes the internal structure of hadrons and the hadronization mechanism. This model has successfully explained dihadron production in unpolarized and single-polarized processes. During the research process, while maintaining the transverse momentum dependence of the hadron pair, we employ the transverse momentum dependent(TMD) factorization framework, using this method and the model, we first simulate the asymmetry in the COMPASS energy region and compare it with experimental data. Furthermore, we predict the same asymmetry at the HERMES, expecting to provide valuable theoretical references for relevant experimental studies. 				
	\end{abstract}
	\maketitle	
	\section{Introduction}
	\label{I}	
The study of dihadron fragmentation functions (DiFFs) is a focal point in both theoretical and experimental aspects. DiFFs describe the probability distribution of the transformation of a quark or an antiquark into two hadrons and everything else through strong interactions. They were first mentioned in Ref.~\cite{Konishi:1979cb}. The evolution equations of DiFFs have been successively examined in Refs.~\cite{Vendramin:1980wz,Vendramin:1981te}, with more in-depth study in Refs.~\cite{DEFLORIAN2004139,Majumder_2007}. Ref.~\cite{Ceccopieri_2007} extended the research perspective by analyzing for the first time the evolution kernel as the functions of the hadron pair invariant mass $M_h$. Ref.~\cite{Collins:1994ax} introduced the transversely polarized dihadron fragmentation function for spin analysis of transversely polarized fragmentation quarks, leading to the definition of $H_1^{\sphericalangle}$. Ref.~\cite{Bianconi_2000} comprehensively probed the leading order distortion in dihadron fragmentation and clarified the definition of the relevant function. Ref.~\cite{Jaffe_1998} started the whole business on dihadron fragmentation to access the quark transversity distributions. Regarding the exploration of nucleon transverse spin, DiFFs~\cite{Collins_1994} have played an important role in the study of nucleon spin structure. Ref.~\cite{Bacchetta_2003} introduced the partial-wave analysis and provided new positive qualitative conditions, offering a clearer understanding of the hadron-pair system. Ref.~\cite{Bacchetta:2003vn} went one step further by extending the analysis to the subleading twist, integrating over the transverse component of hadron pair momentum, and seamlessly connecting to the transverse momentum dependent case. The cross section expression for production of two hadrons in SIDIS within TMD factorization is presented in Ref.~\cite{Gliske:2014wba}. The Collins effect~\cite{Collins:1992kk} and back-to-back dihadron production in $e^+e^-$ annihilation~\cite{Anselmino:2008jk} are commonly used techniques for extracting the chiral-odd transversity distribution. The in-depth study of the dihadron DiFFs has provided additional possibilities for such analyses. Recently, Refs.~\cite{pitonyak2023number,Cocuzza_2024,Cocuzza:2023vqs} presented a series of valuable research results on DiFFs. The transverse spin distributions were initially extracted by relying on the convolution of single hadron SIDIS data $h_1 \otimes H_1^{\perp}$, which can not be separated from the transverse momentum of quarks involved in the chiral-odd Collins fragmentation function. The chiral-odd dihadron fragmentation function $H_1^\sphericalangle$~\cite{,Bacchetta_2003,Radici:2001na} couples with a specific function $h_1$ and plays a crucial role at the leading twist level. The BELLE collaboration's measurements of the azimuthal asymmetry in the distribution of charged pion pairs in annihilation of $e^+e^-$~\cite{Vossen_2011} have motivated the parameterization of $H_1^\sphericalangle$ for up and down quarks~\cite{Courtoy:2012ry}. Recent
studies~\cite{Bacchetta:2011ip,Bacchetta:2012ty,Radici:2015mwa,Radici:2016lam,Radici:2018iag} have extracted $h_1$ from SIDIS and proton-proton collision data. In the mean time, the spectator model~\cite{Bianconi:1999uc,Bacchetta:2006un,Bacchetta:2008wb}, and Nambu-Jona-Lasinio (NJL) quark model~\cite{Matevosyan_2014,Matevosyan_2013,PhysRevLett.120.252001} are used to make predictions for the DiFFs.

In particle physics research, the hadron pair production in SIDIS has attracted much attention, especially with respect to its azimuthal asymmetry. The HERMES collaboration~\cite{HERMES:2008mcr} and COMPASS collaboration~\cite{Adolph:2012nw,Adolph:2014fjw} have begun to measure such asymmetries in the hadron pair production with unpolarized or transversely polarized targets. In addition, the BELLE collaboration has measured the azimuthal asymmetry for the production of back-to-back hadron pair. More recently, the COMPASS collaboration~\cite{Sirtl:2017rhi} has measured the azimuthal asymmetry results for dihadron production using the longitudinally polarized proton target. When the incident lepton beam is unpolarized or longitudinally polarized, one of the many modulations $\sin(3\phi_h-\phi_R)$ modulation appears in our field of view~\cite{Bacchetta_2003}. Other modulations under the same experimental conditions, involving different underlying mechanisms, are discussed in Refs.~\cite{Luo:2019frz,Luo:2020axe,Luo:2020wsg}. Here $\phi_{ h }$ is the azimuthal angle of the hadron pair system, and $\phi_{ R }$ is the angle between the lepton and dihadron plane. Experimentally, COMPASS measurements show that the $\sin(3\phi_h-\phi_R)$ asymmetry is near zero at the current accuracy. Theoretically, in the parton model, this asymmetry arises mainly from the coupling of the function $h_{1L}$ and the T-odd DiFF $H_{1}^{\perp}$.

In this paper, we investigate the $\sin(3\phi_h-\phi_R)$ asymmetry in the SIDIS process for dihadron production. After performing partial waves expansion, this asymmetry is found to arise from the contribution of $h_{1L}H_{1,OT}^{\perp}$ where $H_{1,OT}^{\perp}$ arises from the interference of $s$- and $p$-waves. In the Amsterdam notations, $h_{1L}$ describes a transversely polarized quark in a longitudinally polarized proton, appearing at leading twist. We use the spectator model~\cite{Bacchetta:2006un} to calculate $H_{1,OT}^{\perp}$ and find that the loop contributions have to be considered to obtain a non-vanishing result. Using these results, the $\sin(3\phi_h-\phi_R)$ asymmetry is theoretically predicted at the COMPASS kinematics and compared with the COMPASS preliminary data.

The remainder of the article is structured as follows. The basic framework of the dihadron fragmentation function is introduced in Chapter II. In Chapter III, the calculation of the $\sin(3\phi_h-\phi_R)$ azimuthal asymmetry in the SIDIS process of dihadron production is presented, where an unpolarized lepton beam is scattered with a longitudinally polarized proton target. Chapter IV applies the spectator model to calculate the dihadron fragmentation function $H_{1,OT}^{\perp}$ for T-odd functions. Chapter V provides numerical results for the $\sin(3\phi_h-\phi_R)$ azimuthal asymmetry under the kinematics of the COMPASS and HEMERS measurements. The present work is summarized in Chapter VI.
\section{The basic framework}	
\label{II}
Consider the dihadron fragmentation function process $q\rightarrow \pi^+\pi^-X$, in which a quark with momentum $k$ fragments into two unpolarized pions with masses $M_1$, $M_2$, and momenta $P_1$, $P_2$. For convenience, we introduce two vectors $P_h=(P_1+P_2)$ and $R=(P_1-P_2)/2$, with $P_h$ and $R$ representing the total momentum and the relative momentum of the dihadron. In addition, $M_h$ represents the invariant mass of the dihadron. We describe a four-dimensional vector $\vec{a}$ as $[a^{-},a^{+},\vec{a_T}]$ in terms of the longitudinal light-cone coordinates $ a ^ { \pm } = \frac { a ^ { 0 } \pm a ^ { 3 } } { \sqrt { 2 } }$ and the transversal light-cone coordinates $\vec{a}_T = (a ^ {1},a^ {2})$. We also introduce $x$ which represents the longitudinal momentum fraction of the final state quark. And $z_i$ is the longitudinal component of the hadron $h_i$ found in the fragmentation of the quark. The ratio of the light-cone momentum fraction carried by the hadron pair with respect to the fragmentation quark is defined as $z$. For convenience, the axes are chosen by virtue of the condition $\vec{P}_{hT} = 0$. In order to perform the partial wave expansion of the dihadron fragmentation function, we need to give the following dynamical quantities in the center of mass system of the dihadron. Here the momentum $P_h^\mu$, $k_\mu$ and $R_\mu$ can be written as~\cite{Bacchetta:2006un}
\begin{align}
	P_{h}^{\mu}&= \left[ P_{h}^{-}, \frac{M_{h}^{2}}{2P_{h}^{-}}, \vec{0}_{T}\right],\\k^{\mu}&= \left[ \frac{P_{h}^{-}}{z}, \frac{z(k^{2}+ \vec{k}_{T}^{2})}{2P_{h}^{-}}, \vec{k}_{T}\right],\\R^{\mu}&= \bigg[ - \frac{| \vec{R}|P_{h}^{-}}{M_{h}}\cos \theta , \frac{| \vec{R}|M_{h}}{2P_{h}^{-}}\cos \theta \notag,\\&| \vec{R}| \sin \theta \cos \phi _{R},| \vec{R}| \sin \theta \sin \phi _{R}\bigg]\notag\\&=\left [ - \frac { | \vec{ R } | P _ { h } ^ { - } } { M _ { h } } \cos \theta , \frac { | \vec { R } | M _ { h } } { 2 P _ { h } ^ { - } } \cos \theta , \vec{ R } _ { T } ^ { x } , \vec { R } _ { T } ^ { y }\right ],
\end{align}
where	 	
\begin{align}
		|\vec{R}|&= \sqrt{\frac{M_{h}^{2}}{4}-m_{\pi}^{2}}.
\end{align}

In this physical context, $m_\pi$ represents the mass of the $\pi$ meson and $\theta$ is defined as the polar angle between the direction of $P_1$ in the dihadron center of mass system and the direction of $P_h$ in the laboratory system~\cite{Bacchetta_2003}. With the help of these four momenta, we are able to derive some practical relations
\begin{align}
		&P_{h}\cdot k= \frac{M_{h}^{2}}{2z}+z \frac{k^{2}+ \vec{k}_{T}^{2}}{2},
		\\
		&P_{h}\cdot R=0,
		\\
		&R \cdot k=\bigg(\frac{M_{h}}{2z}-z \frac{k^{2}+ \vec{k}_{T}^{2}}{2M_{h}}\bigg)| \vec{R}| \cos \theta - \vec{k}_{T}\cdot \vec{R}_{T}.
\end{align}
\section{THE $\sin(3\phi_{ h }-\phi_{ R })$ ASYMMETRY OF DIHADRON PRODUCTION IN SIDIS}	
	\label{III}
\begin{figure}[H]
		\centering %图片居中
		\includegraphics[width=0.4\textwidth]{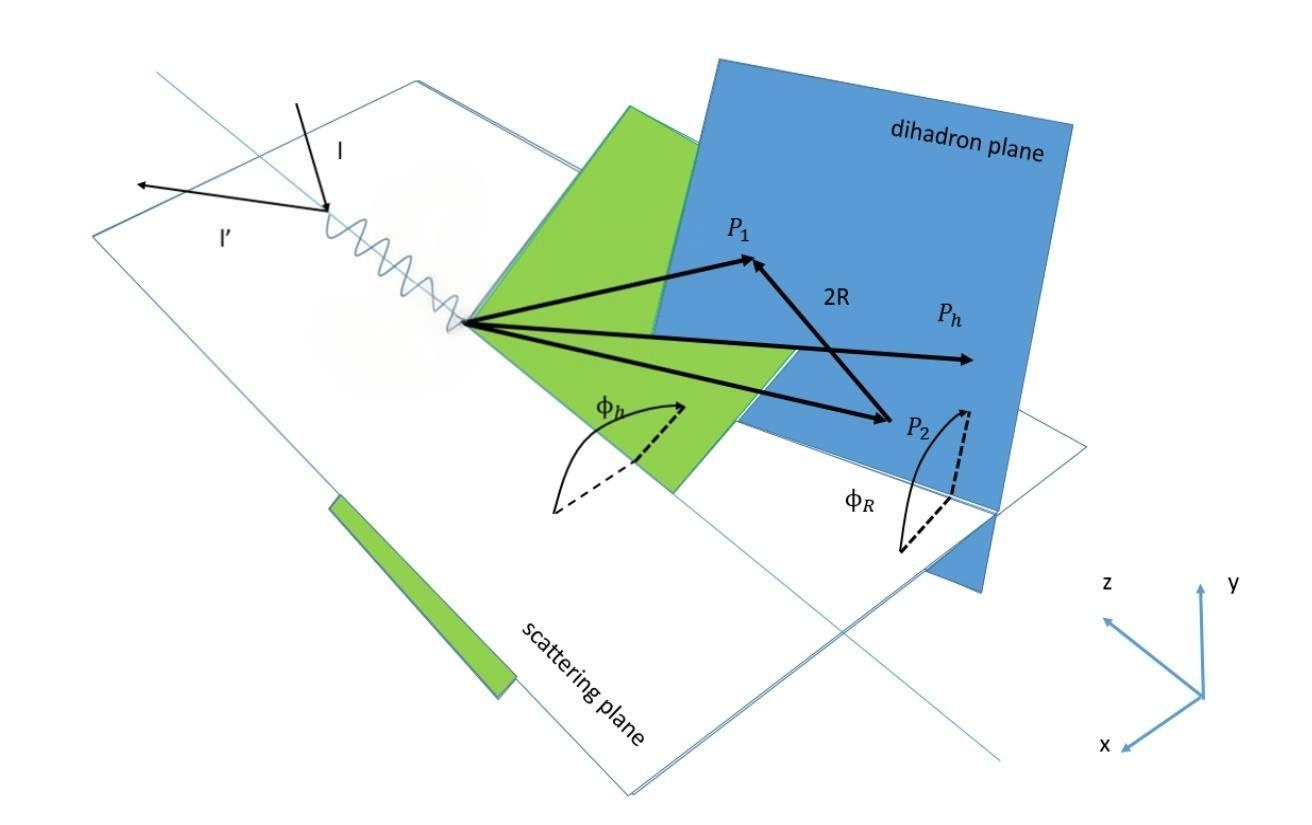} %插入图片，[]中设置图片大小，{}中是图片文件名
		\caption{Angle definitions involved in the measurement of the single longitudinal spin asymmetry in SIDIS production of two
			hadrons.} %最终文档中希望显示的图片标题
		\label{Fig1} %用于文内引用的标签
\end{figure}

The SIDIS process shown in Fig.\Ref{Fig1}, in which an unpolarized lepton $\mu$ collides with a longitudinally polarized nucleon target to produce dihadrons $h^{+}$ and $h^{-}$ where the incoming and outgoing four-momenta of the lepton are denoted as $\ell$ and $\ell^{'}$, and the momentum of the proton target is $P$, with mass $M$ and polarization $S$. In this process, the parton momentum is $p$, and the virtual photon momentum $q=\ell-\ell^{'}$. The final state quark momentum is $k=p+q$, which subsequently fragments up into two unpolarized hadrons with masses $M_1$, $M_2$ and momenta $P_1$, $P_2$, respectively, along with some unobservable states $X$. In order to clarify the relation between the differential cross section of the process and the structure function on which the dihadron fragmentation function depends, we use the following kinematic variables
\begin{align}
			&x = \frac { p ^ { + } } { P ^+ } , y = \frac { P \cdot q } { P \cdot \ell } , z = \frac { P _ { h } ^ { - } } { k ^ { - } } = z _ { 1 } + z _ { 2 } ,\\&z _ { i } = \frac { P _ { i } ^ { - } } { k ^ { - } } , Q ^ { 2 } = - q ^ { 2 } , s = ( P + \ell	 ) ^ { 2 } ,\\&P _ { h } = P _ { 1 } + P _ { 2 } , R = \frac { R _ { 1 } - R _ { 2 } } { 2 } , M _ { h } ^ { 2 } = P _ { h } ^ { 2 } .
\end{align}

The DiFFs $D_1$ and $H_1 ^ {\perp}$ can be extracted from the quark-quark correlation function $\Delta(k; P_h; R)$
\begin{align}
	\Delta( k , P_h , R) =\sum \kern -1.3 em \int_X \ \frac { d ^ { 4 } \xi } { ( 2 \pi ) ^ { 4 } } e ^ { i k\cdot \xi } \langle 0 | \psi ( \xi ) | P _ { h } , R ; X \rangle\notag\\ \langle X ; P _ { h } , R | \overline { \psi } ( 0 ) | 0 \rangle | _ { \xi ^ { - } = \vec { \xi }_T  = 0 }\notag\\= \frac{1}{16\pi} \left\{ D_1\slashed{n}_- + G_1^\perp\gamma_5\frac{\varepsilon_T^{\mu\nu} R_{T\mu} k_{T\nu}}{M_h^2}\slashed{n}_- \right. \notag\\
	\left. + H_1^\perp \frac{\sigma_{\mu\nu} k_T^\mu n_-^\nu}{M_h} + H_{1}^{\sphericalangle} \frac{\sigma_{\mu\nu} R_T^\mu n_-^\nu}{M_h} \right\},
\end{align}
where $n_-=\frac{1}{\sqrt{2}}[1,0,0,-1]$, similarly, we need to represent the quark-quark correlation function in the leading twist form in the center-of-mass system. There is a specific relation between the correlation function under these two different representations, which reads as follows
\begin{align}\label{eq12}
	\Delta ( z , k _ { T } ^ { 2 } , \cos \theta , M _ { h } ^ { 2 } , \phi _ { R } , \phi _ { k } ) = \frac { | \vec { R } | } { 1 6 z M _ { h } } \int d k ^ { + } \Delta ( k , P _ { h } , R ).
\end{align}

By first decomposing the correlation function of the center-of-mass system according to the generally possible Dirac structure, and then projecting the desired scalar function, the following relation can be obtained
\begin{align}
	4 \pi {\rm T r} \left[ \Delta ( z , k _ { T } ^ { 2 } , \cos \theta , M _ { h } ^ { 2 } , \phi _ { R } , \phi _ { k }) i \sigma ^ { \alpha - } \gamma _ { 5 } \right] = \frac { \varepsilon _ { T } ^ { \alpha \beta } k ^\beta _ { T } } { M _ { h } } H _ { 1 } ^ { \perp } + \frac { \varepsilon _ { T } ^ { \alpha \beta } R ^\beta _ { T } } { M _ { h } } H_{1}^{\sphericalangle},
\end{align}
where $i\sigma^{\alpha-}=-\frac{1}{2}(\gamma^\alpha\gamma^--\gamma^-\gamma^\alpha)$, and $\gamma^{-}$ is the negative light-cone Dirac function.

The DiFFs $D_1$ and $H_1^{\perp}$ can be expanded into the partial waves form associated with the corresponding dihadron system~\cite{Bacchetta_2003}. The dependence on $\vec{k}_T \cdot \vec{R}_T$ makes the expansion more complicated 
\begin{widetext}
\begin{align}
	D _ { 1 }(z,k^2_T,\cos \theta,M_h^2,\phi _ { R },\phi _ { k })&= D _ { 1 , OO }(z,M_h^2)+ D _ { 1 , O L }(z,M_h^2) \cos \theta + D _ { 1 , L L }(z,M_h^2) \frac { 1 } { 4 } ( 3 \cos ^ { 2 } \theta - 1 )\notag\\& + \cos ( \phi _ { k } - \phi _ { R } ) \sin \theta ( D _ { 1 , OT }+D_{ 1 , L T }\cos \theta ) + \cos ( 2 \phi _ { k } - 2 \phi _ { R } ) \sin ^ { 2 } \theta D _ { 1 , T T },\\ H _ { 1 } ^ { \perp }(z,k^2_T,\cos \theta,M_h^2,\phi _ { R },\phi _ { k }) &= H _ { 1 , OO }^\perp (z,M_h^2)+ H _ { 1 ,O L } ^ { \perp }(z,M_h^2) \cos \theta + H _ { 1 , L L } ^ { \perp }(z,M_h^2) \frac { 1 } { 4 } ( 3 \cos ^ { 2 } \theta - 1 )\notag\\&+2\cos(\phi _ { k } - \phi _ { R })  \sin \theta ( H _ { 1 , OT }^ { \perp } + H _{1,LT}^ { \perp } \cos \theta ).
\end{align}
\end{widetext}

$H_{1,OT}^{\perp}$ originates from the interference of $s$- and $p$-waves, $H _ { 1 , OO }^ \perp$ is the sum of the pure contributions from the $s$- wave ($\frac{1}{4} H_{1,OO,s}^{\perp}$) and $p$- wave ($\frac{3}{4} H_{1,OO,p}^{\perp}$). We then consider the azimuthal asymmetry of the SIDIS process, in which the unpolarized muon and longitudinally polarized nucleon targets scatter. Within the TMD framework using the defining $A(y)=1-y+\frac{y^2}{2}$, the differential cross section of the process is~\cite{Radici:2001na}
\begin{widetext}
\begin{align}\label{eq16}
		\frac { d ^ { 8 } \sigma _ { U L } } { d x d y d z d \phi _ { h } d \phi _ { R } d \cos \theta d \vec { P } _ { h \bot } ^ { 2 } d M _ { h } ^ { 2 } } = \frac { \alpha ^ { 2 } } { 2 \pi s x y ^ { 2 } } A ( y ) \sum _ { q } e _ { q } ^ { 2 }
		\mathcal{I} \Bigg[ f_1^q\times
		 \Bigg\{
		D_{1,OO}^q
		+ D_{1,OL}^q\cos\theta
		+ \notag\\D_{1,LL}^q \frac{1}{4}(3\cos^2\theta - 1) + \sin\theta \cos(\phi_h - \phi_R)\left( D_{1,OT}^q + D_{1,LT}^q \cos\theta \right)  + \sin^2\theta \cos[2(\phi_h - \phi_R)] \, D_{1,TT}^q
		\Bigg\} \Bigg],
\end{align}
and Ref.~\cite{Bacchetta_2003} presented a complete formula for the longitudinally polarized nucleon, we can get
\begin{align}\label{eq17}
	\frac { d ^ { 8 } \sigma _ { U L } } { d x d y d z d \phi _ { h } d \phi _ { R } d \cos \theta d \vec { P } _ { h \bot } ^ { 2 } d M _ { h } ^ { 2 } } = \frac { \alpha ^ { 2 } } { 2 \pi s x y ^ { 2 } } A ( y ) \sum _ { q } e _ { q } ^ { 2 }\{\sin \theta \sin (3\phi_{h}-\phi_{R}  )\times\notag\\(\mathcal{-I}\left[\frac{4(\vec{p}_T\cdot\hat{P}_{h\perp})(\vec{k}_T\cdot\hat{P}_{h\perp})^2-2(\vec{p}_T\cdot\vec{k}_T)(\vec{k}_T\cdot\hat{P}_{h\perp})-\vec{k}_T^2(\vec{p}_T\cdot\hat{P}_{h\perp})}{2MM_h^2}\right]h^\perp_{1L}(-\frac{2M_h}{|\vec{k}_T|}H_{1,OT}^{\perp}))\}.
\end{align}
\end{widetext}

The structure functions appearing in Eq.~\eqref{eq17} are convolutional forms with specific weights
\begin{align}
	\mathcal{I} \left[ f \right] = \int d ^ { 2 } \vec { p }_T d ^ { 2 } \vec { k } _ { T } \delta ( \vec { p } _ { T } - \vec { k } _ { T } - \frac { \vec{ P }_ { h _ { \bot }  } } { z } ) \left[ f \right].
\end{align}     
where $\phi_{ R }$ and $\phi_{ S }$ are the azimuthal angles of the transverse vectors $\vec{ R }_T$ and $\vec{ S }_T$ with respect to the lepton scattering plane, respectively. The $\hat{P}_{h\perp}$ satisfies $\hat{P}_{h\perp}=\vec{ P}_{h\perp}/|\vec{P}_{h\perp}|$. For convenience the unpolarized or longitudinally polarized states are denoted by the labels $U$ and $L$. 

In Eq.~\eqref{eq16}, $f_{1 }^{q}$ and $D_{1,OO}^{q}$ are the unpolarized PDF and the unpolarized DiFF with flavor $q$, respectively. In Eq.~\eqref{eq17}, $h^\perp_{1L}$ is a distribution function of twist-2 which couples to the T-odd DiFF $H_{1,OT}^{\perp}$. To obtain the asymmetry we need to integrate over the $\cos\theta$ and the result is as follows
\begin{widetext}
\begin{align}\label{eq19}
	A^{\sin(3\phi_h-\phi_R)}_{UL}=\frac{\pi}{4}\frac{\Sigma_qe^2_q\int\mathcal{I}\big{[}-[\frac{4(\vec{p}_T\cdot\hat{P}_{h\perp})(\vec{k}_T\cdot\hat{P}_{h\perp})^2-2(\vec{p}_T\cdot\vec{k}_T)(\vec{k}_T\cdot\hat{P}_{h\perp})-\vec{k}_T^2(\vec{p}_T\cdot\hat{P}_{h\perp})}{2MM_h^2}]h^\perp_{1L}(-\frac{2M_h}{|\vec{k}_T|}H_{1,OT}^{\perp})\big{]}}{\Sigma_qe^2_q\int\mathcal{I}[f_{ 1 }^{ q }D_{1,OO}]}.
\end{align}   
\end{widetext}	

\section{THE MODEL CALCULATION OF $H_{1,OT}^\perp$}
\label{IV}
In this section, we need to calculate the DiFF $H_{1,OT}^{\perp}$ under the spectator model. Theoretically, the absence of imaginary phases leads to a vanishing $H_{1,OT}^{\perp}$ for the tree diagram correlation function. However, in the case of the DiFF, the quark-dihadron interaction vertex $F^{s*}F^p$ is defined in complex form as a way of generating this imaginary phase. Thus, a correlation function similar to the Ref.~\cite{Bacchetta:2006un} can be obtained
\begin{align}\label{eq20}
		\Delta ^ { q } ( k , P _ { h } , R )\notag& = \frac { 1 } { ( 2 \pi ) ^ { 4 } } \frac { ( \slashed { k } + m ) } { ( k ^ { 2 } - m ^ { 2 } ) ^ { 2 } }\\& \times( F ^ { s * } e ^ { - \frac { k ^ { 2 } } { \Lambda _ { s } ^ { 2 } } } + F ^ { p * } e ^ { - \frac { k ^ { 2 } } { \Lambda _ { p } ^ { 2 } } } \slashed { R } )\notag\\& \times( \slashed{k} -\slashed{ P} _ { h } + M _ { s } ) \times ( F ^ { s } e ^ { - \frac { k ^ { 2 } } { \Lambda _ { s } ^ { 2 } } } + F ^ { p } e ^ { - \frac { k ^ { 2 } } { \Lambda _ { P } ^ { 2 } } } \slashed { R } )\notag\\&\times (\slashed{k} + m ) \cdot 2 \pi \delta ( ( k - P _ { h } ) ^ { 2 } - M _ { s } ^ { 2 } ) ,
\end{align}
where $m$ and $M_s$ represent the masses of the fragmented quark as well as the spectator quark, respectively. The $s$-wave and $p$-wave vertex structures $F^s$, $F^p$ are usually written in the following form~\cite{Bacchetta:2006un}
\begin{align}
		&F ^ { s } = f _ { s },\notag \\&F ^ { p } = f _ { \rho } \frac { M _ { h } ^ { 2 } - M _ { \rho } ^ { 2 } - i \Gamma _ { \rho } M _ { \rho } } { ( M _ { h } ^ { 2 } - M _ { \rho } ^ { 2 } ) ^ { 2 } + \Gamma _ { \rho } ^ { 2 } M _ { \rho } ^ { 2 } }+ f _ { \omega } \frac { M _ { h } ^ { 2 } - M _ { \omega } ^ { 2 } - i \Gamma _ { \omega } M _ { \omega } } { ( M _ { h } ^ { 2 } - M _ { \omega } ^ { 2 } ) ^ { 2 } + \Gamma _ { \omega } ^ { 2 } M _ { \omega } ^ { 2 } }  \notag\\&-if_\omega^{\prime}\frac{\sqrt{\lambda(M_\omega^2,M_h^2,m_\pi^2)}\Theta(M_\omega-m_\pi-M_h)}{4\pi\Gamma M_\omega ^2[4M_\omega ^2m_\pi^2+\lambda(M_\omega^2,M_h^2,m_\pi^2)]^{\frac{1}{4}}}.	
\end{align}
where $\lambda(M_\omega^2$,$M_h^2,m_\pi^2)=( M _ { \omega } ^ { 2 } - ( M _ { h } + m _ { \pi } ) ^ { 2 } ) ( M _ { \omega } ^ { 2 } - ( M _ { h } - m _ { \pi } ) ^ { 2 } )$, $\Theta$ denotes the unit step function. The first two terms of $F^p$ can be identified with the contributions of the $\rho$ and the $\omega$ resonances decaying into two pions. The masses and widths of the two resonances are adopted from the PDG~\cite{ParticleDataGroup:2004fcd}: $M_\rho=0.776~$GeV, $\Gamma_\rho=0.150~$GeV, $M_\omega=0.783~$GeV and $\Gamma_\omega=0.008~$GeV.

Substituting Eq.~\eqref{eq20} into Eq.~\eqref{eq12} yields
	\begin{align}\label{eq22}
		\Delta ^ { q } ( &z , k_T^2 , \cos\theta , M_h^2, \phi_R , \phi_k )=\frac{|\vec{R}|}{256\pi^3z(1-z)M_hk^-}\notag\\&\times\bigg[|F^s|^2e ^ { - \frac { 2k ^ { 2 } } { \Lambda _ { s } ^ { 2 } } }\frac{(\slashed{ k}+m)(\slashed{ k}-\slashed{P} _ { h }+M_S)(\slashed{ k}+m)}{(k^2-m^2)^2}\notag\\&+|F^p|^2e ^ { - \frac { 2k ^ { 2 } } { \Lambda _ { p } ^ { 2 } } }\frac{(\slashed{ k}+m)\slashed{ R }(\slashed{ k}-\slashed{P} _ { h }+M_S)\slashed{ R }(\slashed{ k}+m)}{(k^2-m^2)^2}\notag\\&+F^{s*}F^pe ^ { - \frac { 2k ^ { 2 } } { \Lambda _ { sp } ^ { 2 }} }\frac{(\slashed{ k}+m) (\slashed{ k}-\slashed{P} _ { h }+M_S)\slashed{ R }(\slashed{ k}+m)}{(k^2-m^2)^2}\notag\\&+F^{s*}F^pe ^ { - \frac { 2k ^ { 2 } } { \Lambda _ { sp } ^ { 2 }} }\frac{(\slashed{ k}+m)\slashed{ R }(\slashed{ k}-\slashed{P} _ { h }+M_S)(\slashed{ k}+m)}{(k^2-m^2)^2}\bigg].
	\end{align}
Here $z$-dependent $\Lambda$ truncates $\Lambda _ { sp }$ and $\Lambda _ { s, p }$ to satisfy the following relation
	\begin{align}
		\frac{2}{ \Lambda _ { sp } ^ { 2 }}=\frac{1}{ \Lambda _ { s } ^ { 2 }}+\frac{1}{ \Lambda _ { p } ^ { 2 }},
	\end{align}
where $\Lambda _ { s, p }$ satisfies the following structure and $a$,$\beta$ and $\gamma$ are the model parameters that will be given afterwards, and the $k^2$ is fixed by the on-shell condition of the spectator
	\begin{align}
		\Lambda _ { s,p }=a_{s,p}z^{\beta_{s,p}}(1-z)^{\gamma_{s,p}},
	\end{align}
	\begin{align}
		k^2=\frac{z}{1-z}\vec{k_T^2}+\frac{M_s^2}{1-z}+\frac{M_s^2}{z}.
	\end{align}
In Eq.~\eqref{eq22}, the terms of the $|F^s|^2$ and $|F^p|^2$ couplings are pure $s$-wave and $p$-wave contributions, respectively, and therefore they do not contribute to $H_{1,OT}^{\perp}$. $H_{1,OT}^{\perp}$ originates exclusively from the $s$- and $p$-wave interferences, described by the last two terms in Eq.~\eqref{eq22}, where the required imaginary phase originates in the $p$-wave vertex $F^p$. Following the approach of Ref.~\cite{Bacchetta:2006un}, for the tree diagram results of $H_{1, OT}^{\perp}$, the mass of the input quark can be set to be zero GeV. We have verified that even if the input quark is given a small mass, they have essentially no effect on the model predictions of the DiFF and the resulting asymmetry. Therefore, the tree diagram contribution of $H_{1, OT}^{\perp}$ vanishes, and it is then necessary to consider the one-loop diagram contribution.
	
Write the one-loop contribution of the correlation function in the Fig.~\ref{Fig2} according to Feynman rule

\begin{widetext}	
	\begin{align}\label{eq26}
		\Delta_a^q ( z , k _ { T } ^ { 2 } , \cos \theta , M _ { h } ^ { 2 } , \phi _ { R } , \phi_k)&=i\frac { C _ { F } \alpha _ { s } } { 3 2 \pi ^ { 2 } ( 1 - z ) P _ { h } ^ { - } } \cdot \frac { | \vec { R } | } { M _ { h } } \cdot \frac { ( \slashed{k} + m ) } { ( k ^ { 2 } - m ^ { 2 } ) ^ { 3 } } ( F ^ { s * } e ^ { - \frac { k ^ { 2 } } { \Lambda _ { s } ^ { 2 } } } + F ^ { p * } e ^ { - \frac { k ^ { 2 } } { \Lambda _ { p } ^ { 2 } } }  \slashed{ R } ) (\slashed{ k} - \slashed{P} _ { h } + M _ { s } ) \notag\\&\times ( F ^ { s } e ^ { - \frac { k ^ { 2 } } { \Lambda _ { s } ^ { 2 } } } + F ^ { p } e ^ { - \frac { k ^ { 2 } } { \Lambda _ { p } ^ { 2 } } } \slashed{R} ) (\slashed{ k} + m ) \int \frac { d ^ { 4 } \ell } { ( 2 \pi ) ^ { 4 } } \frac { \gamma ^ { \mu } ( \slashed{k} - \slashed{\ell} + m ) \gamma _ { \mu } ( \slashed{k} + m ) } { ( ( k - \ell ) ^ { 2 } - m ^ { 2 } + i \varepsilon ) ( \ell ^ { 2 } + i \varepsilon ) },
	\end{align}
		\begin{align}\label{eq27}
		\Delta _ { b } ^ { q } ( z , k _ { T } ^ { 2 } , \cos \theta , M _ { h } ^ { 2 } , \phi _ { R }, \phi_k)& =i\frac { C _ { F } \alpha _ { s } } { 3 2 \pi ^ { 2 } ( 1 - z ) P _ { h } ^ { - } } \frac { | \vec{ R } | } { M _ { h } } \frac { ( \slashed{k} + m ) } { ( k ^ { 2 } - m ^ { 2 } ) ^ { 2 } } ( F ^ { s * } e ^ { - \frac { k ^ { 2 } } { \Lambda _ { s } ^ { 2 } } } + F ^ { p * } e ^ { - \frac { k ^ { 2 } } { \Lambda _ { p } ^ { 2 } } } { \slashed{R} } ) (\slashed{ k} -\slashed{ P} _ { h } + M _ { s } )\notag\\&\int \frac { d ^ { 4 } \ell } { ( 2 \pi ) ^ { 4 } } \frac { \gamma ^ { \mu } ( \slashed{k} - \slashed{P} _ { h } -\slashed{l} + M _ { s } ) ( F ^ { s } e ^ { - \frac { k ^ { 2 } } { \Lambda _ { s } ^ { 2 } } } + F ^ { p } e ^ { - \frac { k ^ { 2 } } { \Lambda _ { p }^2 }  } {\slashed{ R} } ) ( \slashed{k} -\slashed{\ell}  + m ) \gamma _ { \mu } (\slashed{k} + m ) } { ({ k }-P _ { h } - \ell ) ^ { 2 } - M _ { s } ^ { 2 } + i \varepsilon ) ( ( k - \ell ) ^ { 2 } - m ^ { 2 } + i \varepsilon ) ( \ell ^ { 2 } + i \varepsilon ) },
	\end{align}
\begin{figure}[H] %H为当前位置，!htb为忽略美学标准，htbp为浮动图形
	\centering %图片居中
	\includegraphics[width=0.6\textwidth]{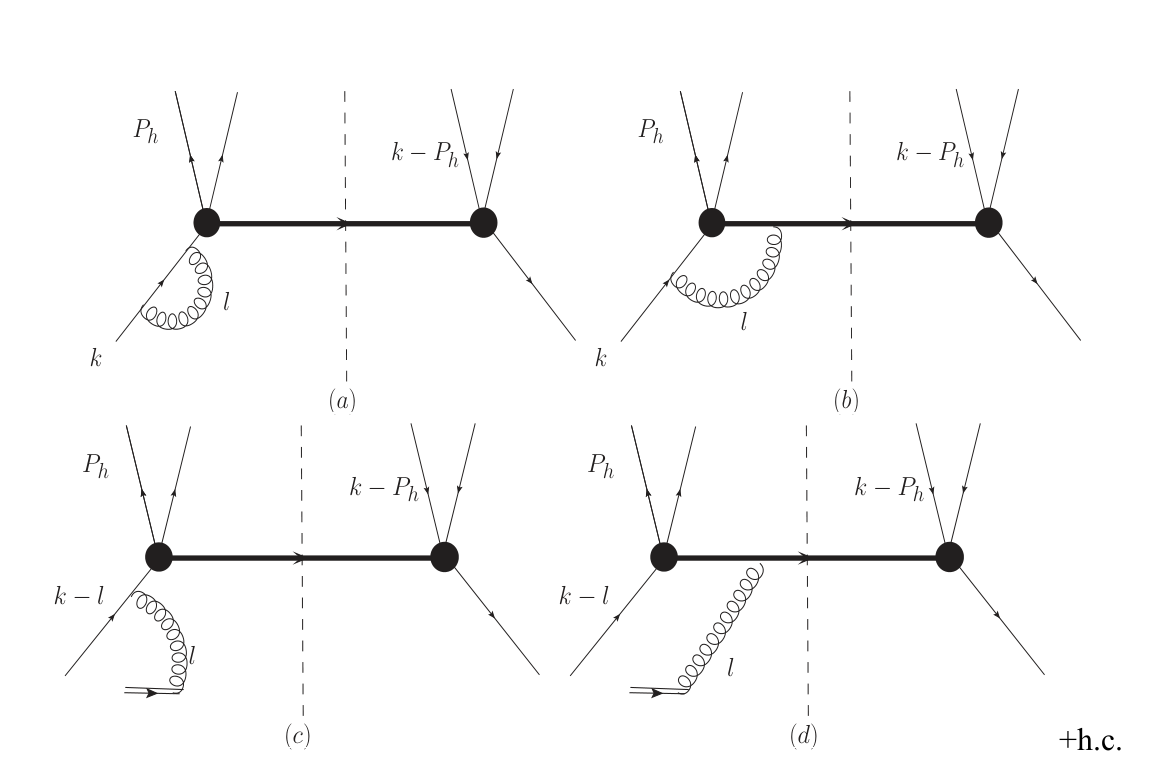} %插入图片，[]中设置图片大小，{}中是图片文件名
	\caption{ One loop order corrections to the fragmentation function of a quark into a meson pair in the spectator model. Where	h.c. represents the hermitian conjugations of these diagrams.} %最终文档中希望显示的图片标题
	\label{Fig2} %用于文内引用的标签
\end{figure}	
	\begin{align}\label{eq28}
		\Delta_c^q( z , k _ { T } ^ { 2 } , \cos \theta , M _ { h } ^ { 2 } , \phi _ { R }, \phi_k)&=i\frac { C _ { F } \alpha _ { s } } { 3 2 \pi ^ { 2 } ( 1 - z ) P _ { h } ^ { - } } \cdot \frac { | \vec{ R } | } { M _ { h } } \cdot \frac { (\slashed{ k} + m ) } { ( k ^ { 2 } - m ^ { 2 } ) ^ { 2 } } ( F ^ { s * } e ^ { - \frac { k ^ { 2 } } { \Lambda _ { s } ^ { 2 } } } + F ^ { p * } e ^ { - \frac { k ^ { 2 } } { \Lambda _ { p } ^ { 2 } } } \slashed{ R } ) (\slashed{ k} -\slashed{ P} _ { h } + M _ { s } )\notag\\& \times ( F ^ { s } e ^ { - \frac { k ^ { 2 } } { \Lambda _ { s } ^ { 2 } } } + F ^ { p } e ^ { - \frac { k ^ { 2 } } { \Lambda _ { p } ^ { 2 } } } \slashed{ R } ) \int \frac { d ^ { 4 } \ell } { ( 2 \pi ) ^ { 4 } } \frac { ( \slashed{k} + m ) \gamma ^ { - } ( \slashed{k} - \slashed{\ell} + m ) } { ( ( k - \ell ) ^ { 2 } - m ^ { 2 } + i \varepsilon ) ( - \ell^-  \pm i \varepsilon ) ( \ell^ { 2 } + i \varepsilon ) },
	\end{align}
	
	\begin{align}\label{eq29}
		\Delta _ { d } ^ { q } ( z , k _ { T } ^ { 2 } , \cos \theta , M _ { h } ^ { 2 } , \phi _ { R } , \phi _ {k})& = i \frac { C _ { F } \alpha _ { s } } { 3 2 \pi ^ { 2 } ( 1 - z ) P _ { h } ^ { - } } \cdot \frac { | \vec{ R } | } { M _ { h } } \cdot \frac { ( \slashed{k} + m ) } { k ^ { 2 } - m ^ { 2 } } ( F ^ { s * } e ^ { - \frac { k ^ { 2 } } { \Lambda _ { s } ^ { 2 } } } + F ^ { p * } e ^ { - \frac { k ^ { 2 } } { \Lambda_ { p } ^ { 2 } } } \slashed{ R } ) (\slashed {k} - \slashed{P} _ { h } + M _ { s } )\notag\\&\int \frac { d ^ { 4 } \ell } { ( 2 \pi ) ^ { 4 } } \frac{\gamma^-(\slashed{k}-\slashed{P}_h-\slashed{\ell}+m_s)( ( F ^ { s } e ^ { - \frac { k ^ { 2 } } { \Lambda _ { s } ^ { 2 } } } + F ^ { p  } e ^ { - \frac { k ^ { 2 } } { \Lambda_ { p } ^ { 2 } } } \slashed{ R } ) )(\slashed{k}-\slashed{\ell}+m)}{ ({ k }-P _ { h } - \ell) ^ { 2 } - M _ { s } ^ { 2 } + i \varepsilon ) ( ( k - \ell) ^ { 2 } - m ^ { 2 } + i \varepsilon )(-\ell^-\pm i\varepsilon) ( \ell ^ { 2 } + i \varepsilon )}.
	\end{align}
\end{widetext}	

In Eqs. (\ref{eq26}--\ref{eq29}), we use the Feynman rule $1/(-\ell^{-}\pm i\xi)$ for the eikonal propagator, and this Feynman rule is also applicable to the vertex formed between the eikonal line and the gluon. As a matter of principle, the Gaussian form factors in these formulas should depend on the loop momentum $\ell$. In order to simplify the integrals, we follow the choice made in Ref.~\cite{Marcel:2021hrv}, where the dependence on $\ell$ is discarded and only these Gaussian form factors are assumed to have a $k^2$ dependence. Similar choices are made in Ref.~\cite{Bacchetta:2002tk,Bacchetta:2003xn,Amrath:2005gv}, which also lead to reasonable final results.

Here we apply the Cutkosky cutting rules
	\begin{align}
		\frac { 1 } {\ell ^ { 2 } + i \varepsilon } \rightarrow - 2 \pi i \delta ( \ell ^ { 2 } ) , \frac { 1 } { ( k - \ell ) ^ { 2 } + i \varepsilon } \rightarrow - 2 \pi i \delta ( ( k - \ell ) ^ { 2 } ) .      
	\end{align}
	
Using the above convention, the final result of $H_{1,OT}^\perp$ is obtained:
\begin{align}
		H_{1,OT}^{\perp a}&=0,\\
		H_{1,OT}^{\perp b}&=\frac{1}{2\pi^3}\Bigg{[}\frac{C_F\alpha_s|\vec{R}|^2}{1-z}\cdot|F^{s*}F^p|e^\frac{-2k^2}{\Lambda^2_{sp}}\Bigg{]}\notag\\&\times\frac{1}{(k^2-m^2)^2}k_TC_b,\\
		H_{1,OT}^{\perp c}&=0,\\                                                             
		H_{1,OT}^{\perp d}&=-\frac{1}{2\pi^3}\Bigg{[}\frac{C_F\alpha_s|\vec{R}|^2}{1-z}\cdot|F^{s*}F^p|e^\frac{-2k^2}{\Lambda^2_{sp}}\Bigg{]}\notag\\&\times\frac{1}{(k^2-m^2)}((I_2-\mathcal{A})k_T),
\end{align}
with
	\begin{align}
		C_b=\mathcal{A}(2k^2+2m^2+2mM_s)+\mathcal{B}(2m^2+2M_h^2-2M_s^2)\notag\\+\mathcal{A}_0(-k^2-m^2)+\mathcal{B}_0(M_s^2-m^2-M_h^2)+I_2(m^2-k^2).
	\end{align}
The coefficients $\mathcal{A}$ and $\mathcal{B}$ denote the following functions
	\begin{align}
		\mathcal{A} &= \frac { I _ { 1 } } { \lambda ( k , M _ { h } , M _ { s } ) } [ 2 k ^ { 2 } ( k ^ { 2 } - M _ { s } ^ { 2 } - M _ { h } ^ { 2 } ) \frac { I _ { 2 } } { \pi }\notag\\& + ( k ^ { 2 } + M _ { h } ^ { 2 } - M _ { s } ^ { 2 } ) ] ,\\ \mathcal{B} &= - \frac { 2 k ^ { 2 } } { \lambda ( k , M _ { h } , M _ { s } ) } I _ { 1 } \left(1 + \frac { k ^ { 2 } + M _ { s } ^ { 2 } - M _ { h } ^ { 2 } } { \pi } I _ { 2 } \right),
	\end{align}
which originate from the decomposition of the following integral~\cite{Lu:2015wja}
	\begin{align}
		\int d ^ { 4 } \ell \frac { \ell ^ { \mu } \delta ( \ell^ { 2 } ) \delta \left[ ( k - \ell ) ^ { 2 } - m ^ { 2 } \right] } { ( k - P _ { h } - \ell ) ^ { 2 } - M _ { s } ^ { 2 } } = \mathcal{A} k ^ { \mu } + \mathcal{B} P _ { h } ^ { \mu }.
	\end{align} 
The functions $I_i$ represent the results of the following integrals
	\begin{align}
		I _ { 1 } &= \int d ^ { 4 } \ell \delta ( \ell ^ { 2 } ) \delta \left[ ( k - \ell ) ^ { 2 } - m ^ { 2 } \right] = \frac { \pi } { 2 k ^ { 2 } } ( k ^ { 2 } - m ^ { 2 } ), \\I _ { 2 } &= \int d ^ { 4 } \ell \frac { \delta ( \ell^ { 2 } ) \delta \left[ ( k - \ell ) ^ { 2 } - m ^ { 2 } \right] } { ( k -\ell - P _ { h } ) ^ { 2 } - M _ { s } ^ { 2 } } = \frac { \pi } { 2 \sqrt { \lambda ( k , M _ { h } , M _ { s } ) } } \notag\\&\ln \bigg( 1 - \frac { 2 \sqrt { \lambda ( k , M _ { h } , M _ { s } ) } } { k ^ { 2 } - M _ { h } ^ { 2 } + M _ { s } ^ { 2 } + \sqrt { \lambda ( k , M _ { h } , M _ { s } ) } } \bigg),
	\end{align}
where $\lambda(k,M_h,M_s) = [k^2 - (M_h + M_s)^2][k^2 - (M_h - M_s)^2]$. Furthermore, we have to calculate the following integrand 
	\begin{widetext}
	\begin{align}
		\int d^4\ell\frac{\ell^\mu\ell^\nu\delta(\ell^2)\delta((k - \ell)^2 - m^2)}{(k - p - \ell)^2 - M_s^2} = k^\mu k^\nu\mathcal{A}_0 + k^\mu p^\nu\mathcal{B}_0 + p^\mu k^\nu\mathcal{C}_0 + p^\mu p^\nu\mathcal{D}_0 + g^{\mu\nu}\mathcal{E}_0,
	\end{align}
where the coefficients $\mathcal{A}_0$, $\mathcal{B}_0$, $\mathcal{C}_0$, $\mathcal{D}_0$ and $\mathcal{E}_0$ are defined as:
\begin{align}
		\mathcal{A}_0 = \frac{(k^2 - m^2)(\mathcal{A}k^4 - \mathcal{B}k^4 - 4\mathcal{A}k^2M_h^2 - 2\mathcal{B}k^2M_h^2 + \mathcal{A}M_h^4 + \mathcal{B}M_h^4 - 2\mathcal{A}k^2M_s^2 + 2\mathcal{B}k^2M_s^2 - 2\mathcal{A}M_h^2M_s^2 + \mathcal{A}M_s^4 - \mathcal{B}M_s^4)}{2k^2(k^4 - 2k^2M_h^2 - 2k^2M_s^2 + M_h^4 - 2M_h^2M_s^2 + M_s^4)},
	\end{align}
	\begin{align}
		\mathcal{B}_0=\mathcal{C}_0=\frac{1}{2}\frac{(k^{2}-m^{2})(\mathcal{A}k^{2}+3\mathcal{B}k^{2}+\mathcal{A}M_{h}^{2}-\mathcal{B}M_{h}^{2}-\mathcal{A}M_{s}^{2}-3\mathcal{B}M_{s}^{2})}{k^{4}-2k^{2}M_{h}^{2}-2k^{2}M_{s}^{2}+M_{h}^{4}-2M_{h}^{2}M_{s}^{2}+M_{s}^{4}},
	\end{align}
\begin{align}
	\mathcal{D}_0=-\frac{(k^{2}-m^{2})(\mathcal{A}k^{2}+2\mathcal{B}k^{2}-\mathcal{B}M_{h}^{2}+\mathcal{B}M_{s}^{2})}{k^{4}-2k^{2}M_{h}^{2}-2k^{2}M_{s}^{2}+M_{h}^{4}-2M_{h}^{2}M_{s}^{2}+M_{s}^{4}},
\end{align}
\begin{align}
	\mathcal{E}_0=-\frac{1}{4}(k^2-m^2)(\mathcal{A}-\mathcal{B}),
\end{align}
and $\mathcal{B}_0$ = $\mathcal{C}_0$ because of the symmetry of the integrand under the exchange of the Lorentz indices $\mu$ and $\nu$.
	\end{widetext}

\section{NUMERICAL RESULTS}
\label{V}		
To determine the parameters of the spectator model, the authors of Ref.~\cite{Bacchetta:2006un} conduct a comparison between the model and the output of the PYTHIA event generator~\cite{Sjostrand:2000wi} that is employed for the HERMES experiment. The parameter values acquired through the fitting process are as follows: $ a_{s}=2.60~\mathrm{GeV}$, $\beta _ { s } =-0 . 7 5 1$, $\gamma _ { s } = - 0 . 1 9 3 $, $a _ { p } = 7 . 0 7~\mathrm{GeV}$, $\beta _ { p } = - 0 . 0 3 8 $, $\gamma _ { p } = - 0 . 0 8 5 $, $f_s=1197~\mathrm{GeV^{-1}}$, $f _ { \rho } = 9 3 . 5 $, $f _ { \omega } = 0 . 6 3 $, $f _ { \omega } ^ { \prime } = 7 5. 2 $, $M _ { s } = 2 . 9 7~M_h$. In this study, as chosen in Ref.~\cite{Bacchetta:2006un}, the input quark mass is set to zero $\mathrm{GeV}$. It is worth emphasizing that the model parameters are obtained by comparing the theoretical predictions with simulations generated from PYTHIA events under HERMES kinematics. Subsequent predictions of the asymmetry will be made under the kinematic conditions of COMPASS and EIC. Since this prediction is based on model parameters obtained under HERMES dynamics, there is necessarily a degree of uncertainty in these parameters. However, in the current work, this uncertainty is neglected and the strong coupling is chosen to be $\alpha_{ s }\approx0.3$.

\begin{figure*}[htbp]
	\centering % 图片居中
	\begin{minipage}{0.45\textwidth} % 左图
		\centering
		\includegraphics[width=\linewidth]{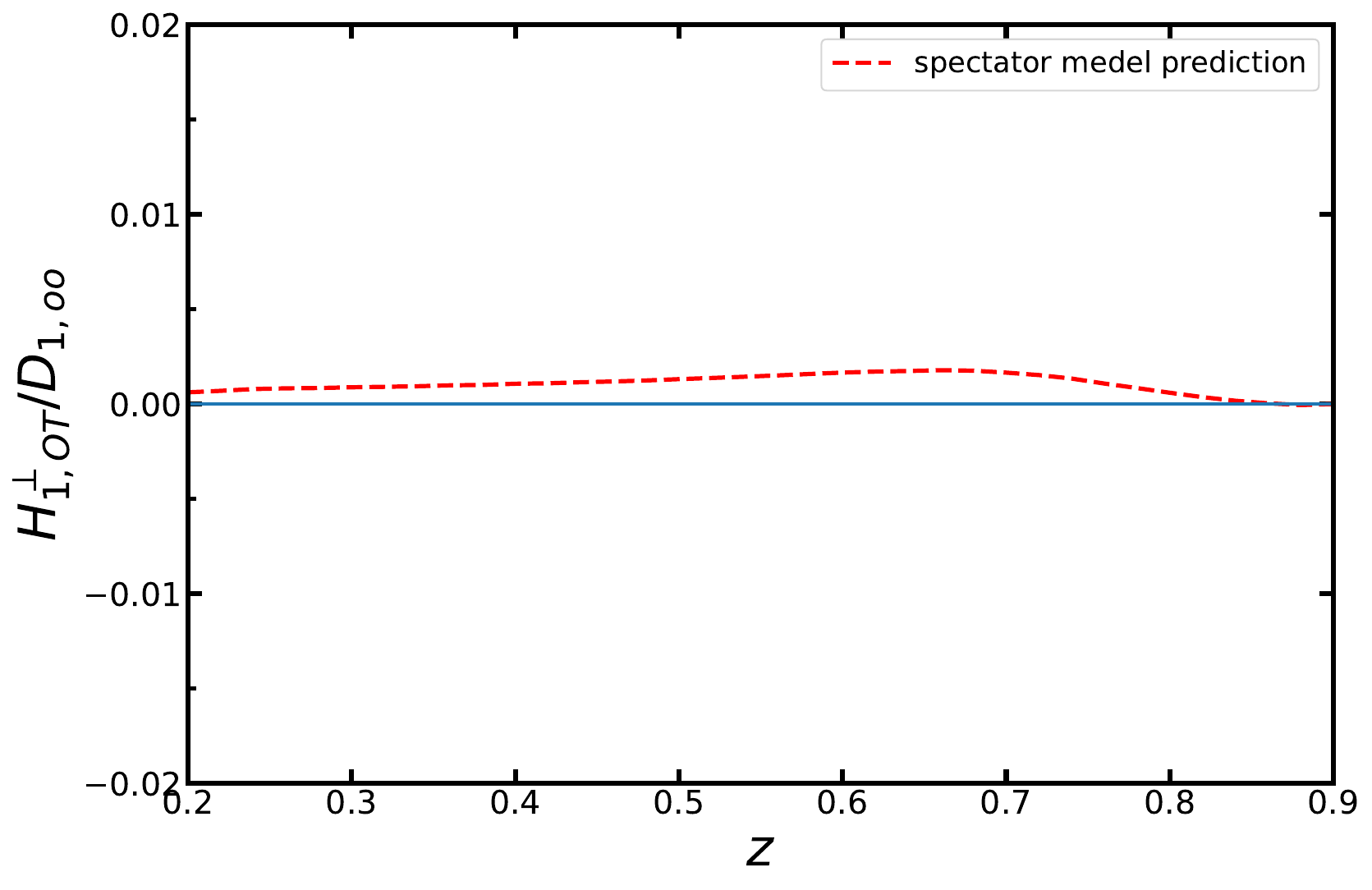} % 图片宽度与 minipage 一致
	\end{minipage}
	\hfill % 增加水平间距
	\begin{minipage}{0.45\textwidth} % 右图
		\centering
		\includegraphics[width=\linewidth]{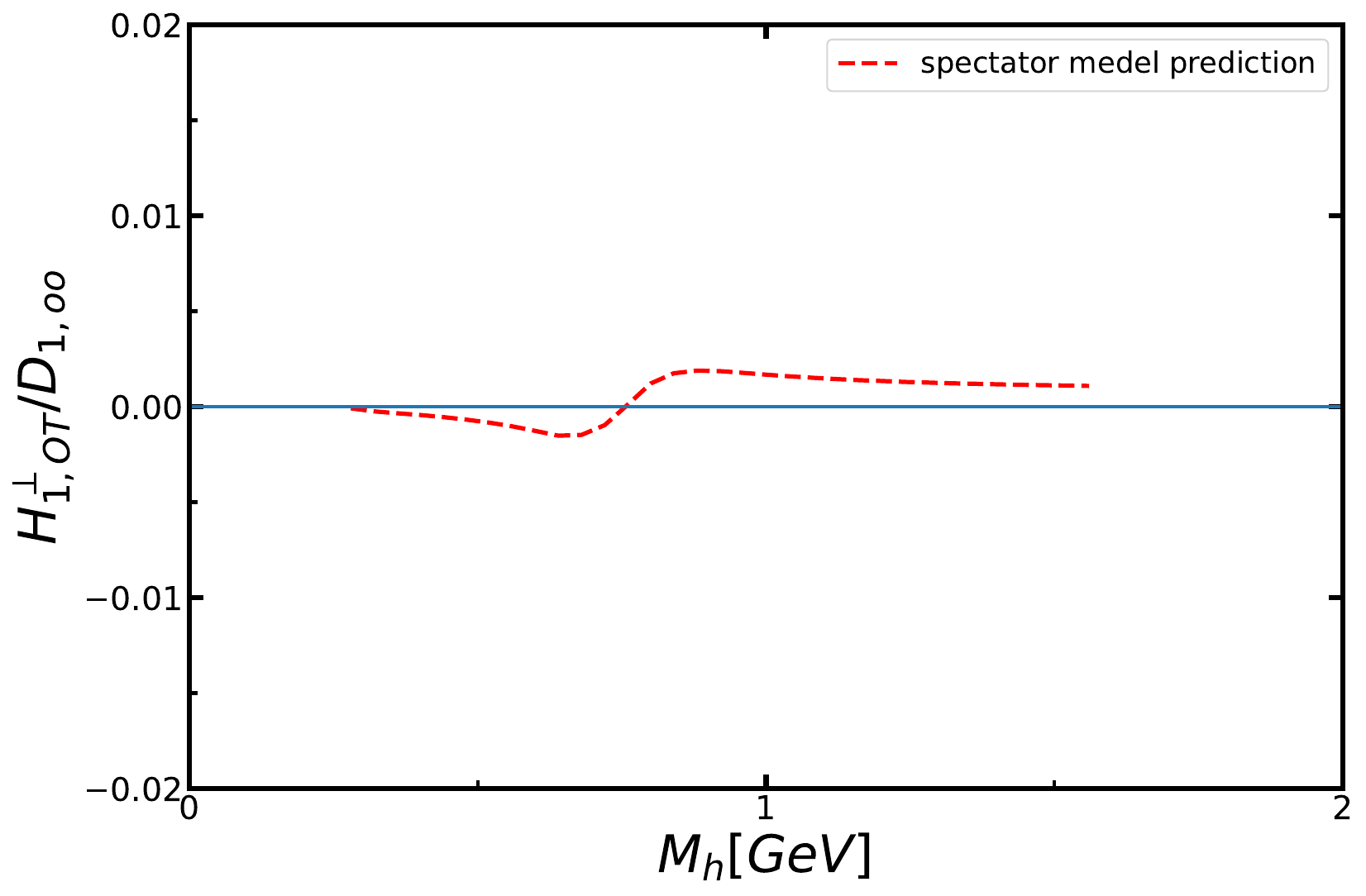} % 图片宽度与 minipage 一致
	\end{minipage}
	\caption{The DiFF $H_{1,OT}^\perp$ as functions of $z$ (left panel) and $M_h$ (right panel) in the spectator model, normalized by the unpolarized DiFF $D_{1,OO}$.} % 图片标题左对齐
	\label{Fig3} % 图片标签
\end{figure*}

In the left and right panels of Fig.~\ref{Fig3}, we plot the ratio of $H_{1,OT}^{\perp}$ relative to $D_{1,OO}$ as a function of $z$ and $M_h$, obtained after integrating over the azimuthal angle $\phi_k$ and the magnitude $k_T$ of the transverse momentum. For the $z$-dependent plot (left panel), we integrate over $M_h$ in the range $0.3~\mathrm{GeV} < M_h < 1.6~\mathrm{GeV}$; conversely, for the $M_h$-dependent plot (right panel), we integrate over $z$ in the range $0.2~\mathrm{GeV}< z < 0.9~\mathrm{GeV}$. The comparison shows that $H_{1,OT}^{\perp}$ is numerically three orders of magnitude smaller compared to the unpolarized dihadron fragmentation function $D_{1,OO}$. At the same time, it is found that there exists a zero point at $M_h = 0.74~\mathrm{GeV}$.

In the following, we present the numerical results for the azimuthal asymmetry of $\sin(3\phi_h-\phi_R)$ in the SIDIS process. This process is defined as the scattering between an unpolarized muon and a longitudinally polarized nucleon target. Based on the principle of isospin symmetry, we find that the fragmentation correlators associated with the processes of $u \rightarrow \pi^+\pi^-X$, $\bar{d} \rightarrow \pi^+\pi^-X$, $d \rightarrow \pi^-\pi^+X$, and $\bar{u} \rightarrow \pi^-\pi^+X$ are similar to one another. Regarding vector transformation, when the sign reversal operation is applied to $\vec{R}$, this operation is mathematically equivalent to performing the angular transformations $\theta\to\pi-\theta$ and $\phi\to\phi+\pi$ respectively. The processes originating from $d \rightarrow \pi^-\pi^+X$ and $\bar{u} \rightarrow \pi^-\pi^+X$, which show a linear dependence on $\vec{R}$ for the DiFF $H_{1,OT}^{\perp}$, produce a negative sign compared to the process originating from $u \rightarrow \pi^+\pi^-X$. When expanding the flavor sum in the numerator of Eq.~\eqref{eq19}, we utilize the isospin symmetry on the DiFF $H_1^{\perp}$. In addition, the sea quark distributions should be generated by the evolution of perturbative QCD and they are zero at the model scale. In the present work, we decide to ignore the QCD evolution which causes the antiquark PDFs $f_1$ and $h_{1L}$ to take zero values. The expressions for the distribution function of $(3\phi_h - \phi_R)$ azimuthal asymmetry with respect to the independent variables $x$, $z$, and $M_h$ below can be accurately derived using Eq.~\eqref{eq19}.

\begin{figure*}[htbp]
	\centering % 图片居中
	\begin{subfigure}[b]{0.3\textwidth}
		\includegraphics[width=\textwidth]{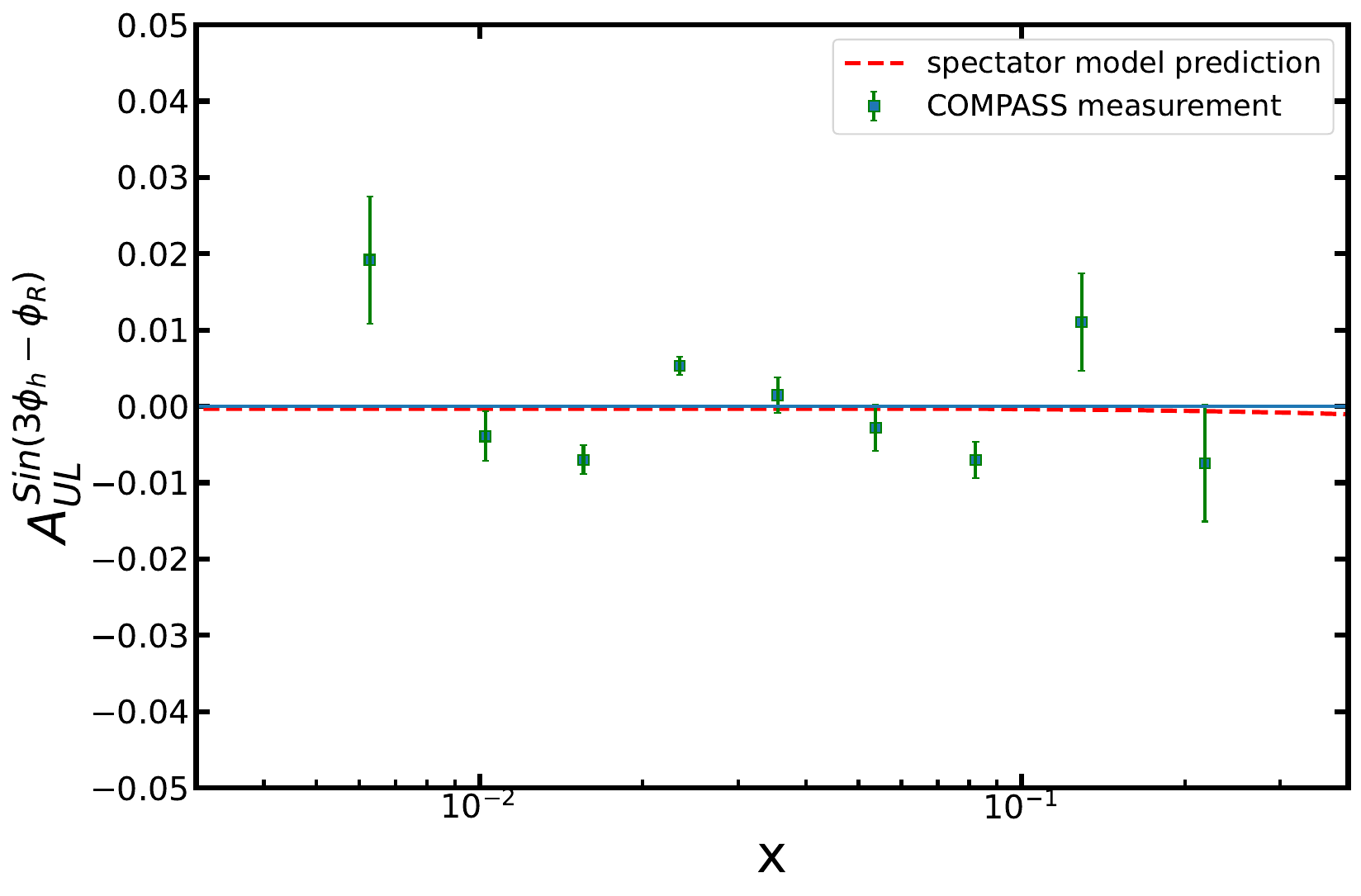}
		\caption{}
		\label{Fig4a}
	\end{subfigure}
	\hfill % 增加水平间距
	\begin{subfigure}[b]{0.3\textwidth}
		\includegraphics[width=\textwidth]{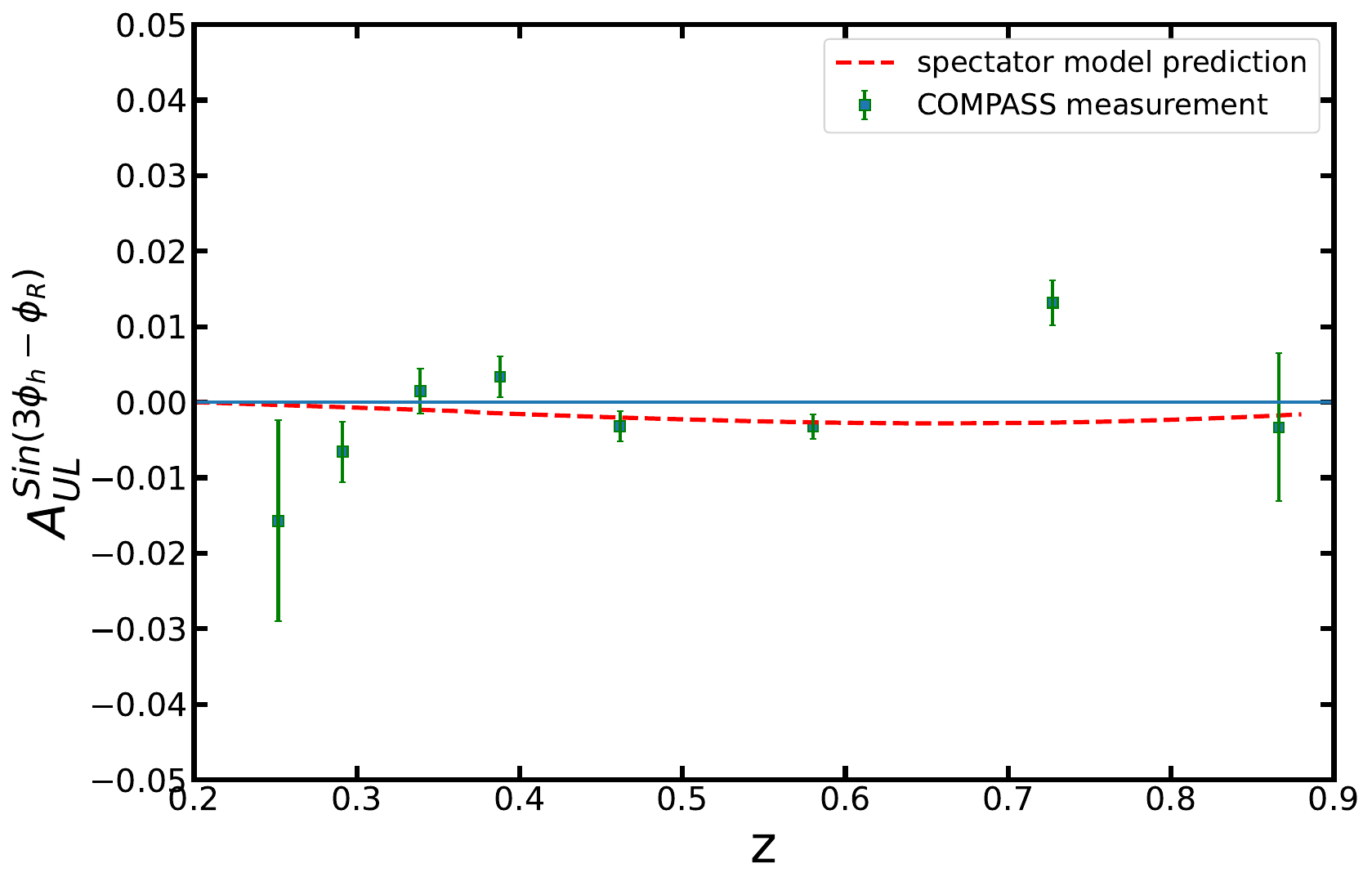}
		\caption{}
		\label{Fig4b}
	\end{subfigure}
	\hfill % 增加水平间距
	\begin{subfigure}[b]{0.3\textwidth}
		\includegraphics[width=\textwidth]{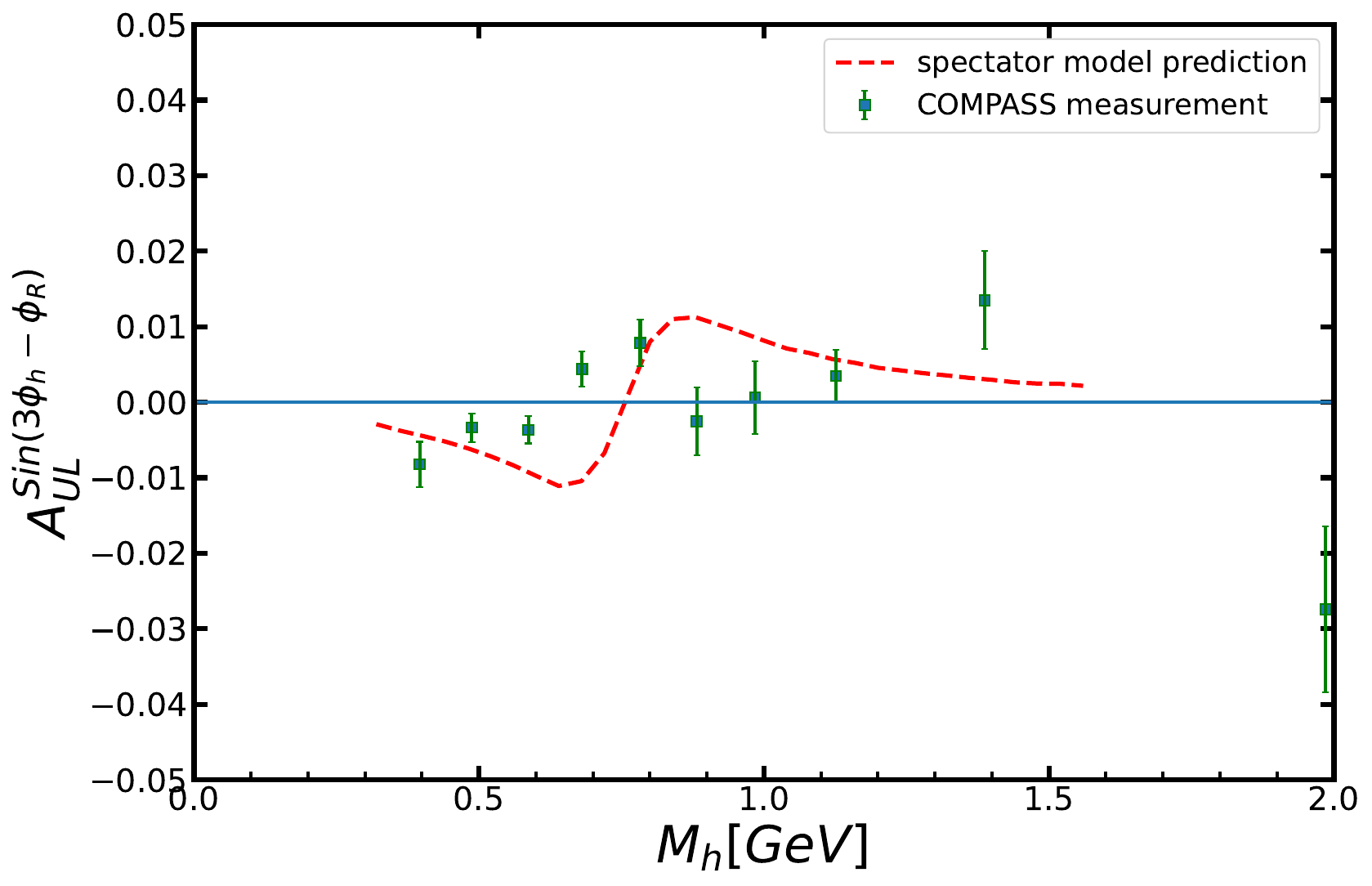}
		\caption{}
		\label{Fig4c}
	\end{subfigure}
	\caption{The $\sin(3\phi_h-\phi_{ R })$ azimuthal asymmetry in the SIDIS process of unpolarized muons off longitudinally polarized nucleon target as a functions of $x$ (Fig.~\ref{Fig4a}), $z$ (Fig.~\ref{Fig4b}) and $M_h$ (Fig.~\ref{Fig4c}) at COMPASS. The full circles with error bars show the preliminary COMPASS data~\cite{Bacchetta:2006un} for comparison. The dashed curves denote the model prediction.} % 图片标题左对齐
	\label{Fig4}
\end{figure*}

\begin{widetext}
	\begin{align}
		A^{\sin(3\phi_h-\phi_R)}_{UL}(x)=\bigg(\int dydz 2M_hdM_hd \cos\theta d^2\vec{k_T}d^2\vec{p_T}d^2\vec{P}_{h\perp}(\frac{k_Tp_T\pi}{4})\delta(\vec{p}_T-\vec{k}_T-\frac{\vec{P}_{h\perp}}{z})(-\notag\\\left[\frac{4(\vec{p}_T\cdot\vec{P}_{h\perp})(\vec{k}_T\cdot\vec{P}_{h\perp})^2-2|\vec{P}_{h\perp}|^2(\vec{p}_T\cdot\vec{k}_T)(\vec{k}_T\cdot\vec{P}_{h\perp})-|\vec{P}_{h\perp}|^2\vec{k}_T^2(\vec{p}_T\cdot\vec{P}_{h\perp})}{2|\vec{P}_{h\perp}|^3MM_h^2}\right](4h^{\perp u}_{1L}-h^{\perp d}_{1L})(-\frac{2M_h}{|\vec{k}_T|}H_{1,OT}^{\perp})\bigg)\big{/}\notag\\\bigg(\int dydz 2M_hdM_hd \cos\theta d^2\vec{k_T}d^2\vec{p_T}d^2\vec{P}_{h\perp}\delta(\vec{p}_T-\vec{k}_T-\frac{\vec{P}_{h\perp}}{z})(4f_{1}^{u}(p_{T}^{2})+f_{1}^{d}(p_{T}^{2}))D_{1,OO}(k_{T}^{2})\bigg),\\
		A^{\sin(3\phi_h-\phi_R)}_{UL}(z)=\bigg(\int dxdy 2M_hdM_hd \cos\theta d^2\vec{k_T}d^2\vec{p_T}d^2\vec{P}_{h\perp}(\frac{k_Tp_T\pi}{4})\delta(\vec{p}_T-\vec{k}_T-\frac{\vec{P}_{h\perp}}{z})(-\notag\\\left[\frac{4(\vec{p}_T\cdot\hat{P}_{h\perp})(\vec{k}_T\cdot\hat{P}_{h\perp})^2-2|\vec{P}_{h\perp}|^2(\vec{p}_T\cdot\vec{k}_T)(\vec{k}_T\cdot\vec{P}_{h\perp})-|\vec{P}_{h\perp}|^2\vec{k}_T^2(\vec{p}_T\cdot\vec{P}_{h\perp})}{2|\vec{P}_{h\perp}|^3MM_h^2}\right](4h^{\perp u}_{1L}-h^{\perp d}_{1L})(-\frac{2M_h}{|\vec{k}_T|}H_{1,OT}^{\perp})\bigg)\big{/}\notag\\\bigg(\int dxdy 2M_hdM_hd \cos\theta d^2\vec{k_T}d^2\vec{p_T}d^2\vec{P}_{h\perp}\delta(\vec{p}_T-\vec{k}_T-\frac{\vec{P}_{h\perp}}{z})(4f_{1}^{u}(p_{T}^{2})+f_{1}^{d}(p_{T}^{2}))D_{1,OO}(k_{T}^{2})\bigg),\\
		A^{\sin(3\phi_h-\phi_R)}_{UL}(M_h)=\bigg(\int dxdydz 2M_hd \cos\theta d^2\vec{k_T}d^2\vec{p_T}d^2\vec{P}_{h\perp}(\frac{k_Tp_T\pi}{4})\delta(\vec{p}_T-\vec{k}_T-\frac{\vec{P}_{h\perp}}{z})(-\notag\\\left[\frac{4(\vec{p}_T\cdot\hat{P}_{h\perp})(\vec{k}_T\cdot\hat{P}_{h\perp})^2-2|\vec{P}_{h\perp}|^2(\vec{p}_T\cdot\vec{k}_T)(\vec{k}_T\cdot\vec{P}_{h\perp})-|\vec{P}_{h\perp}|^2\vec{k}_T^2(\vec{p}_T\cdot\vec{P}_{h\perp})}{2|\vec{P}_{h\perp}|^3MM_h^2}\right](4h^{\perp u}_{1L}-h^{\perp d}_{1L})(-\frac{2M_h}{|\vec{k}_T|}H_{1,OT}^{\perp})\bigg)\big{/}\notag\\\bigg(\int dxdydz 2M_hd \cos\theta d^2\vec{k_T}d^2\vec{p_T}d^2\vec{P}_{h\perp}\delta(\vec{p}_T-\vec{k}_T-\frac{\vec{P}_{h\perp}}{z})(4f_{1}^{u}(p_{T}^{2})+f_{1}^{d}(p_{T}^{2}))D_{1,OO}(k_{T}^{2})\bigg),
	\end{align}
\end{widetext}
where the results for $D_{1,OO}$ can be obtained with the help of a similar tree diagram order calculation

	\begin{align}
		&D _ { 1 , O O } ( z , \vec { k } _ { T } ^ { 2 } , M _ { h } ) = \frac { 4 \pi | \vec { R } | } { 2 5 6 \pi ^ { 3 } M _ { h } z ( 1 - z ) ( k ^ { 2 } - m ^ { 2 } ) ^ { 2 } } \notag\\&\Bigg\{ 4 | F ^ { s } | ^ { 2 } e ^ { - \frac { 2 k ^ { 2 } } { \Lambda _ { s } ^ { 2 } } } ( z k ^ { 2 } - M _ { h } ^ { 2 } - m ^ { 2 } z + m ^ { 2 } + 2 m M _ { s } + M _ { s }^ { 2 })  - \notag\\&4 | F ^ { p } | ^ { 2 } e ^ { - \frac { 2 k ^ { 2 } } { \Lambda _ { p } ^ { 2 } } } | \vec{ R } | ^ { 2 } ( - z k ^ { 2 } + M _ { h } ^ { 2 } + m ^ { 2 } ( z - 1 ) + 2 m M _ { s } - M _ { s } ^ { 2 } )\notag\\&+ \frac { 4 } { 3 } | F ^ { p } | ^ { 2 } e ^ { - \frac { 2 k ^ { 2 } } { \Lambda _ { P } ^ { 2 } } } | \vec{ R } | ^ { 2 }\bigg[ 4 \bigg( \frac { M _ { h } } { 2 z } - z \frac { k ^ { 2 } + \vec { k } _ { T } ^ { 2 } } { 2 M _ { h } } \bigg)^ { 2 }\notag\\& + 2 z \frac { k ^ { 2 } - m ^ { 2 } } { M _ { h } } \bigg( \frac { M _ { h } } { 2 z } - z \frac { k ^ { 2 } + \vec{ k } _ { T } ^ { 2 } } { 2 M _ { h } }\bigg )\bigg]\Bigg\}.
	\end{align}
\begin{figure*}[htbp] % H为当前位置，!htb为忽略美学标准，htbp为浮动图形
	\centering % 图片居中
	% 插入子图
	\begin{subfigure}[b]{0.3\textwidth}
		\includegraphics[width=\textwidth]{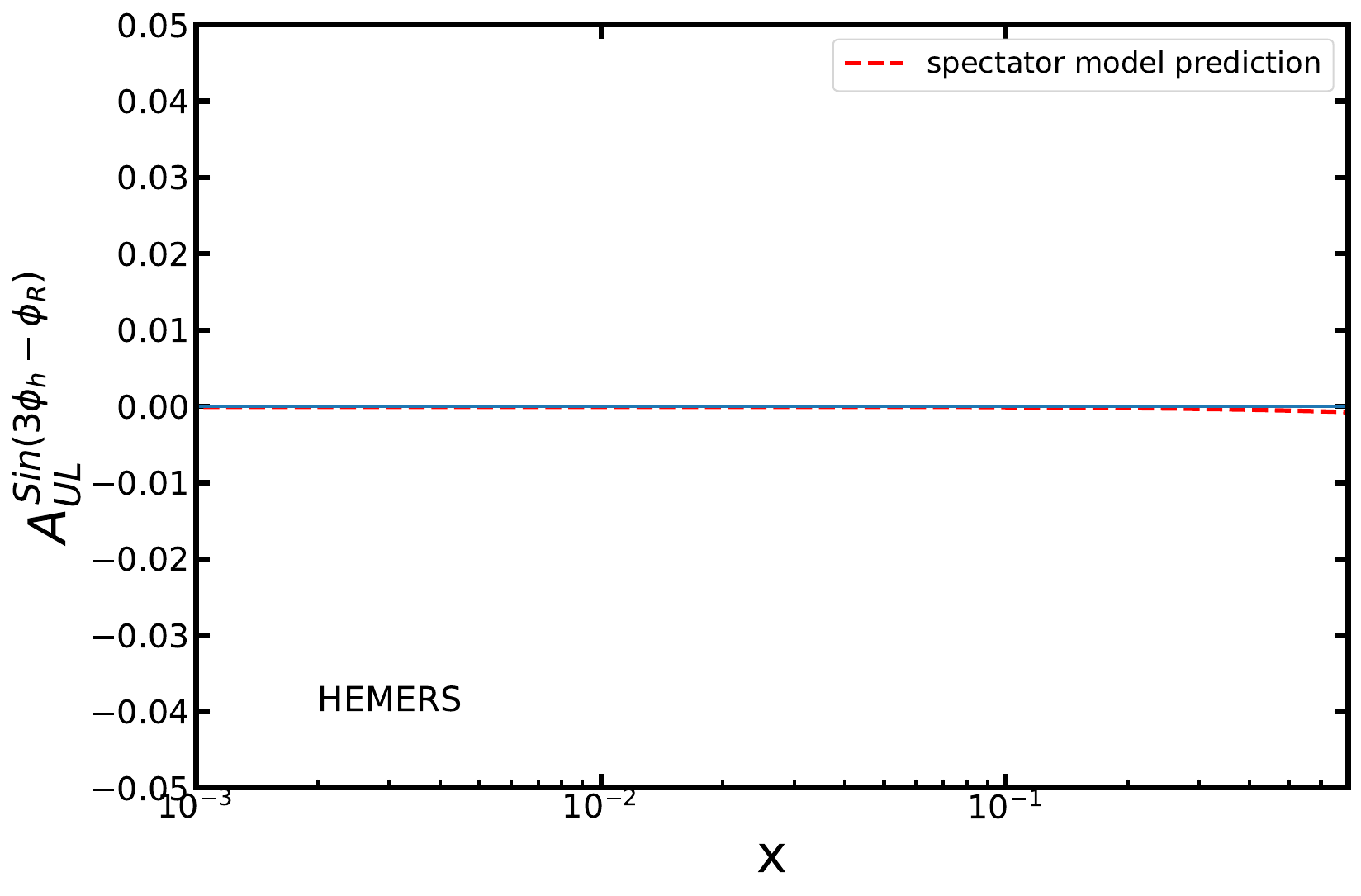} % 插入第一张图片
		\caption{} % 可以为每个子图添加单独的标题
		\label{Fig5a}
	\end{subfigure} \hfill
	\begin{subfigure}[b]{0.3\textwidth}
		\includegraphics[width=\textwidth]{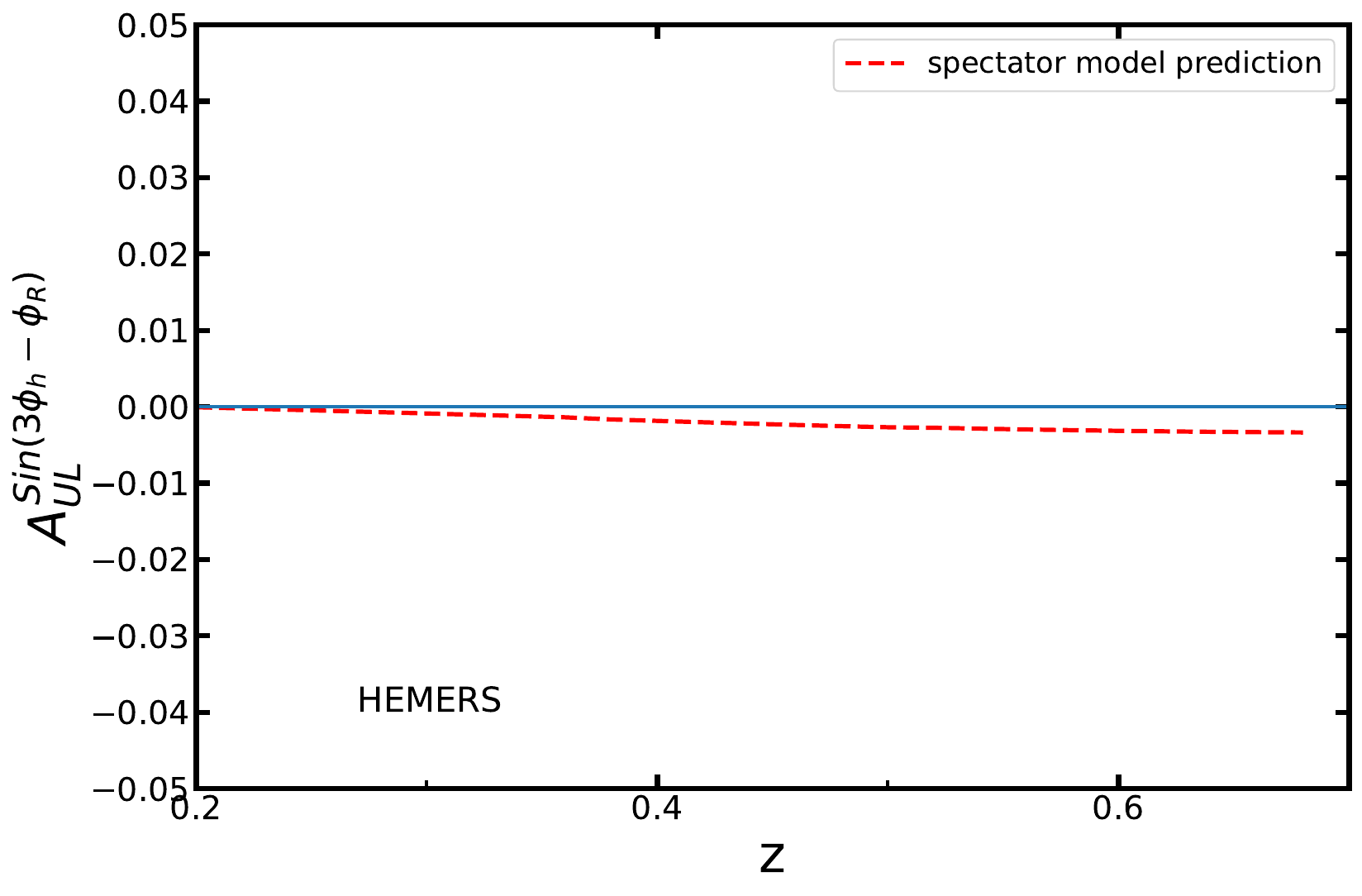} % 插入第二张图片
		\caption{} % 可以为每个子图添加单独的标题
		\label{Fig5b}
	\end{subfigure} \hfill
	\begin{subfigure}[b]{0.3\textwidth}
		\includegraphics[width=\textwidth]{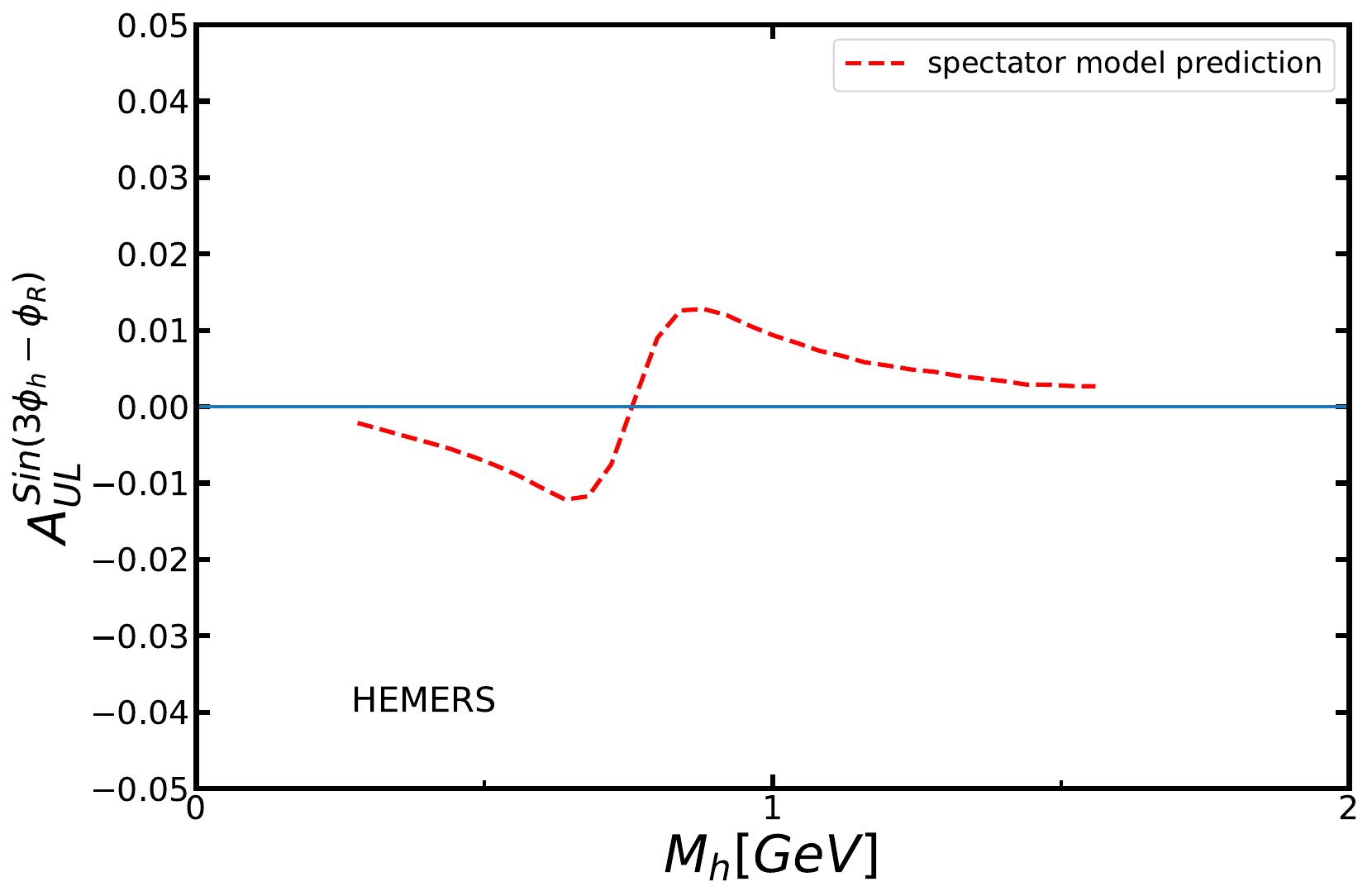} % 插入第三张图片
		\caption{} % 可以为每个子图添加单独的标题
		\label{Fig5c}
	\end{subfigure}
	% 总标题
	\caption{The $\sin(3\phi_h-\phi_{ R })$ azimuthal asymmetry in the SIDIS process of unpolarized muons off longitudinally polarized nucleon target as a functions of $x$ (Fig.~\ref{Fig5a}), $z$ (Fig.~\ref{Fig5b}) and $M_h$ (Fig.~\ref{Fig5c}) at the HERMES($\sqrt{s}=7.2$ GeV).}
	\label{Fig5} % 用于文内引用的标签
\end{figure*}

Regarding the PDFs $f_1$ and $h_{1L}^\perp$, we use the same spectator model results~\cite{Bacchetta:2008af} for uniformity. In order to calculate the value of the $\sin(3\phi_h-\phi_R)$ asymmetry in the process of SIDIS produced by dihadron under the COMPASS kinematics, the following kinematics truncation is used~\cite{Sirtl:2017rhi}
	\begin{itemize}
		\item Cut1 at the COMPASS: $\sqrt{s} = 17.4\ \text{GeV}$, $0.003 < x < 0.4$, $0.1 < y < 0.9$, $0.2 < z < 0.9$, $0.3\ \text{GeV}< M_h < 1.6\ \text{GeV}$, $Q^2 > 1\ \text{GeV}^2$, $W > 5\ \text{GeV}$.
	\end{itemize}
here $W$ is the invariant mass of photon-nucleon system with $W^2=(P+q)^2\approx\frac{1-x}{x}Q^2$. 

After performing numerical operations, under the COMPASS dynamics conditions, our main results are presented in Fig.~\ref{Fig4}. This figure shows both the model predictions and experimental measurements of the $\sin(3\phi_h - \phi_R)$ azimuthal asymmetry. The distributions with respect to $x$ (Fig.~\ref{Fig4a}), $z$ (Fig.~\ref{Fig4b}), and $M_h$ (Fig.~\ref{Fig4c}) are depicted respectively. The dashed line represents our model predictions, while the solid circles with error bars represent the preliminary data from the COMPASS collaboration. Through comparison, we observe that the model predictions effectively characterize the preliminary COMPASS data, which are consistent with zero. Based on the model calculations for $H_{1,OT}^{\perp}$, its relatively small value might be the primary cause of this small magnitude asymmetry.

In addition, for further comparison, we have predicted the same asymmetry on HERMES using the following kinematics truncations. The $x$, $z$, and $M_h$ dependent asymmetries are plotted in Fig.~\ref{Fig5}. As shown, we believe that the overall trend of asymmetry for HERMES is similar to that of COMPASS. The size of the asymmetry is slightly smaller than the size at COMPASS and can still be commensurate with zero in the kinematics of HERMES.
	\begin{itemize}
		\item Cut2~\cite{PhysRevLett.103.152002} at the HERMES: $\sqrt{s} = 7.2\ \text{GeV}$, $0.023 < x < 0.4$, $0.1 < y < 0.95$, $0.2 < z < 0.7$, $0.3\ \text{GeV}< M_h < 1.6\ \text{GeV}$, $Q^2 > 1\ \text{GeV}^2$, $W^2 > 10\ \text{GeV}^2$,
	\end{itemize}
	
	\section{CONCLUSION}
	\label{VI}
In this research, we investigate the single spin asymmetry with a $\sin(3\phi_h-\phi_R)$ modulation in the dihadron production process of SIDIS. With the spectator model results for $D_{1,OO}$, we calculate the T-odd DiFF $H_{1,OT}^{\perp}$ by considering loop contributions to obtain a nonvanishing $H_{1,OT}^{\perp}$. We present the prediction for the $\sin(3\phi_h-\phi_R)$, comparing the prediction with the COMPASS measurements by using the numerical results of the DiFFs and PDFs. The results describe the vanishing data in the COMPASS measurements very well. In addition, we obtain particularly small asymmetries in the dynamics conditions of HERMES. Under the predictions of COMPASS and HERMES, the $M_h$ distribution of the asymmetry exhibits a distinct characteristic structure. The result provides us important clues to get an in-depth understanding of the interference mechanism between $s-$ and $p$- waves during the hadronization process.
	
	\section{Acknowledgments}
	\label{VII}
This work is supported in part by the National Natural Science Foundation of China under Grants No. 12205002, in part by the the Natural Science Foundation of Anhui Province (2108085MA20,2208085MA10), and in part by the key Research Foundation of Education Ministry of Anhui Province of China (KJ2021A0061).

\bibliography{art}	
\end{document}